\documentclass[preprint]{aastex631}

\newcommand       \g            {\,{\rm g}}
\newcommand       \K            {\,{\rm K}}

\newcommand       \s            {\,{\rm s}}

\newcommand       \simali       {\sim\,}

\usepackage{amsmath}

\begin{document}

\title{Optical and Near-Infrared Spectroscopy of the Outbursting Comet 12P/Pons-Brooks}

\correspondingauthor{Ruining Zhao}
\email{rnzhao@nao.cas.cn}

\author[0000-0003-4936-4959]{Ruining Zhao}
\affiliation{South America Center for Astronomy, National Astronomical Observatories, Chinese Academy of Sciences, Beijing 100101, China}
\affiliation{Instituto de Estudios Astrofísicos, Facultad de Ingeniería y Ciencias, Universidad Diego Portales, Santiago, Chile}
\affiliation{School of Astronomy and Space Science, University of Chinese Academy of Sciences, Beijing 100049, China}
\author[0000-0002-5033-9593]{Bin Yang}
\affiliation{Instituto de Estudios Astrofísicos, Facultad de Ingeniería y Ciencias, Universidad Diego Portales, Santiago, Chile}
\affiliation{Planetary Science Institute, 1700 E Fort Lowell Rd STE 106, Tucson, AZ 85719, United States}

\author[0000-0002-6702-7676]{Michael S. P. Kelley}
\altaffiliation{Visiting Astronomer at the Infrared Telescope Facility, which is operated by the University of Hawaii under contract NNH14CK55B with the National Aeronautics and Space Administration.}
\affiliation{Department of Astronomy, University of Maryland, College Park, MD 20742-0001, USA}

\author[0000-0001-8541-8550]{Silvia Protopapa}
  \altaffiliation{Visiting Astronomer at the Infrared Telescope Facility, which is operated by the University of Hawaii under contract NNH14CK55B with the National Aeronautics and Space Administration.}
  \affiliation{Southwest Research Institute, Boulder, CO 80302, USA}

\author[0000-0002-1119-642X]{Aigen Li}
\affiliation{Department of Physics and Astronomy, University of Missouri, Columbia, MO 65211, USA}

\author[0000-0003-3250-2876]{Yang Huang}
\affiliation{School of Astronomy and Space Science, University of Chinese Academy of Sciences, Beijing 100049, China}

\author[0000-0002-2874-2706]{Jifeng Liu}
\affiliation{New Cornerstone Science Laboratory, National Astronomical Observatories, Chinese Academy of Sciences, Beijing 100101, China}
\affiliation{School of Astronomy and Space Sciences, University of Chinese Academy of Sciences, Beijing 100049, China}
\affiliation{Institute for Frontiers in Astronomy and Astrophysics, Beijing Normal University, Beijing, 102206, China}



\begin{abstract}

We present optical and near-infrared (NIR) observations of the outbursting, Halley-type comet 12P/Pons-Brooks. Three NIR spectra were obtained during two outbursts in October and November 2023, with the 3-meter Infrared Telescope Facility and the Palomar 200-inch Telescope, respectively. The NIR spectra exhibited absorption features at 1.5 and 2.0\,${\rm \mu}$m, consistent with the diagnostic absorption bands of water ice, superimposed on a red dust-scattering continuum. We find that the absorption bands and the red continuum can be well explained by micrometer-sized crystalline ice at 140--170\,K, along with sub-micrometer-sized refractory grains (e.g., amorphous carbon). In addition, an optical spectrum was obtained with the Lijiang 2.4-meter Telescope during the November 2023 outburst, which exhibited the emission bands of gaseous CN, C$_3$, C$_2$ and NH$_2$. The C$_3$/CN and C$_2$/CN ratios suggest that 12P/Pons-Brooks was ``typical'' in C$_3$ abundance but somewhat depleted in C$_2$. 
The specific kinetic energy of the 2023 November outburst is estimated to be $\sim8\times10^3$\,J\,kg$^{-1}$, suggesting a likely triggering mechanism similar to 332P/Ikeya--Murakami and 17P/Holmes, i.e., the crystallization of amorphous water ice. A refractory-to-ice ratio of $\simali$1.7--3.2 is derived from the total mass loss of dust and gas, aligning with the lower-end estimates for 67P/Churyumov-Gerasimenko and 1P/Halley. This suggests either a less evolved nucleus or an outburst region enriched in icy materials relative to the bulk nucleus.

\end{abstract}



\section{Introduction} \label{sec:intro}

A cometary outburst is a sudden, unexpected brightening event that typically increases a comet's brightness by $\simali$2--3 magnitudes, with rare cases exceeding $\simali$7.5 magnitudes \citep[or a factor of $\simali$1000;][]{Buzzi:2007,Vales:2010}. The proposed triggering mechanisms for cometary outbursts include cliff collapse \citep{Pajola:2017}, pressure pockets \citep{Muller:2025}, or impacts \citep{Guliev:2022,A'Hearn:2005}, although the detailed processes remain poorly understood. Traditionally, for recurring and large-scale outbursts, the main mechanism to trigger such an activity has been suggested to be the crystallization of amorphous water ice \citep{Smoluchowski:1981, Prialnik:1992b}. Amorphous water ice, formed in the interstellar medium and the outer solar nebula, is considered a major constituent of cometary nuclei \citep{Prialnik:2024}. The (exothermic) crystallization of amorphous water ice leads to the release of trapped gases which results in cometary outbursts. It is also widely proposed as the main mechanism for distant cometary activity \citep{Prialnik2004}. In recent years, it has been recognized that recurring cometary outbursts are a complex phenomenon likely driven by multiple mechanisms operating under different physical conditions. Comet 29P/Schwassmann–Wachmann 1, the most outburst-prone comet, had some of its activities driven by CO and CO$_2$ gas sublimation and the crystallization of amorphous ice \citep{Wierzchos:2020,Lisse:2022}. Cometary outbursts could also result from dust-driven processes such as pressure buildup beneath a dust crust, crust rupture, and rotational modulation of active regions \citep{Trigo-Rodriguez:2010,Miles:2016}. The outbursts of 67P/Churyumov-Gerasimenko have been attributed to two distinct mechanisms: cliff collapses that expose near-surface water ice and pressure-driven release of subsurface volatile pockets, primarily CO$_2$, each associated with different surface topography and outgassing behavior \citep{Muller:2024}.

Amorphous and crystalline ices can be distinguished by subtle differences in the shapes of their absorption bands near 1.5, 2.0, and 3.0\,${\rm \mu}$m. More importantly, crystalline ice shows a narrow, diagnostic absorption feature at 1.65\,${\rm \mu}$m, superimposed on a broad absorption band with sharper edges centered at 1.5\,${\rm \mu}$m. In contrast, amorphous water ice exhibits a smoother profile, a slightly blueshifted band center, and a very weak, less peaked 1.65\,${\rm \mu}$m feature. At higher ice temperatures, the 1.65\,${\rm \mu}$m feature weakens significantly and disappears when the ice temperature exceeds $\sim$\,200\,K \citep{Grundy1998,2008Icar..197..307M}. 

Through near infrared (NIR) spectroscopy, water ice grains have been detected in three outbursting comets: 17P/Holmes, P/2010 H2 (Vales), and 29P/Schwassmann-Wachmann 1. For 17P/Holmes, three spectra were obtained within 7 days after the spectacular outburst in October 2007 at a heliocentric distance ($r_{\rm h}$) of $\sim$2.4\,au, all consistently exhibiting strong 2.0 and 3.0\,${\rm \mu}$m absorption bands \citep{2009AJ....137.4538Y}. However, the entire 1.5\,${\rm \mu}$m band, including the diagnostic feature of crystalline ice at 1.65\,${\rm \mu}$m, was absent, making it difficult to determine the physical structure of the water ice. However, in the case of P/2010 H2 (Vales), the 1.65\,${\rm \mu}$m feature was clearly detected in the two spectra taken 6 and 8 days after a remarkable explosion by 7.5 magnitudes at $r_{\rm h}\sim3.1$\,au, strongly indicating the presence of crystalline ice \citep{2010DPS....42.0509Y}. For 29P/Schwassmann-Wachmann 1, NIR spectra obtained 5 days after a series of outbursts at $r_{\rm h}\sim5.9$ au revealed a weak 2.0\,${\rm \mu}$m absorption feature with a depth of $2.6\pm0.3$\%, tentatively attributed to water ice. However, the 1.5\,${\rm \mu}$m band was not detected, likely due to signal-to-noise limitations, and only an upper limit of $\sim$1\% was derived for the band depth \citep{Protopapa2021ATel}.

The detection of sub-micrometer-sized ice grains 
in 67P/Churyumov-Gerasimenko has also been made through the Alice ultraviolet spectrograph on the {\it Rosetta} spacecraft during an outburst at 3.32\,au through a characteristic electronic absorption edge below 170\,nm \citep{Agarwal:2017}. Notably, the measured dust velocities exceeded those explicable by free water ice sublimation alone, which implies contributions from a pressurized subsurface gas reservoir or energy released during crystallization of amorphous water ice \citep{Agarwal:2017}. 

In addition, the possible presence of water ice grains in the ejecta of 67P/Churyumov-Gerasimenko during its outbursts was also implied by observations made with the {\it Visible InfraRed Thermal Imaging Spectrometer} (VIRTIS) on the {\it Rosetta} spacecraft. During 67P's two outbursts at 1.3\,au, the {\it Rosetta}/VIRTIS observed bright ejecta grains with high bolometric albedos ($\simali$0.7), which were interpreted as either silicate grains (formed through thermal degradation of the carbonaceous material) or icy grains \citep{Bockelee-Morvan:2017}.

Nevertheless, no water ice features have been detected in other outbursting comets, including comet 243P/NEAT, observed only four days after its outburst at a heliocentric distance $r_{\rm h}$ similar to previous detections \citep{Kelley:submitted}. Due to the rarity and substantial differences in water ice detections among comets, consistently characterizing the properties of water ice grains in cometary outbursts remains challenging. Prompt observations of additional outbursting comets are needed.

Comet 12P/Pons-Brooks (hereafter 12P) is a Halley-type comet with a 71-year orbital period. During its current apparition, 12P was recovered in 2020 at $r_{\rm h}=11.89$\,au, showing a broad tail 3$\arcsec$ long, suggesting that its activity began at an even greater distance from the Sun \citep{Ye:2020}. On 2023 July 19, an outburst was observed at $r_{\rm h}=3.89$\,au, with a brightness increase of $\simali$5\,mag \citep{BAA2024}. Since then, recurrent outbursts, both small and large, of up to 5\,mag have been reported, especially from October to December 2023 \citep{2023ATel16270....1U,2023ATel16338....1J,2024ATel16408....1J}.  Interestingly, during its previous apparitions in 1883–1884 and 1954–1955, several outbursts of 3--4 magnitudes were recorded at heliocentric distances ranging from $\gtrsim 3$\,au to 0.79\,au \citep{Beyer:1958,Manzini:2023}. Such repeated activities, occurring decades to centuries apart, share some similarities with that of 17P/Holmes.

Unlike in previous apparitions, the outbursts of this intriguing comet can now be studied using state-of-the-art techniques. We have conducted rapid follow-up observations of 12P in the optical and NIR, soon after the major outbursts, to characterize the gas and dust in its coma and search for water ice grains. By analyzing its water ice features and chemical composition, in this work we explore the mechanisms driving its recurring outbursts and compare its observed properties with those of other notable outbursting comets. The observation and data reduction are described in \S\ref{sec:obs}. The reduced spectra are modeled and analyzed in \S\ref{sec:res}. The results are discussed in \S\ref{sec:dis} and summarized in \S\ref{sec:sum}.


\section{Observations and Data reductions} \label{sec:obs}

In total, three NIR spectra of 12P were obtained with the SpeX spectrograph on the 3-meter Infrared Telescope Facility (IRTF) and the Triple Spectrograph (TSpec) on the Palomar 200-inch Hale Telescope (P200). The SpeX observation was conducted on UT 2023 October 9, about 4 days after the outburst on October 5. The high-throughput prism mode was used with a $0\farcs8\times60\arcsec$ slit, which provided a resolving power of $R=\lambda/\Delta\lambda\sim75$ over 0.7--2.5\,$\mu$m \citep{2003PASP..115..362R}. The comet was observed using in-slit nodding mode with a nodding distance of 30$\arcsec$, and the slit was aligned along the parallactic angle. During the run, two nearby G-type stars from \citet{2020AJ....160..130L} were observed as both telluric correction standards and solar analogs. Internal flats and arcs were taken immediately after the comet and standard star observations using the same instrument settings. The SpeX data were reduced using the method described by \cite{2018ApJ...862L..16P}. 

 The two TSpec observations were conducted on UT November 2 and 3, about 1 and 2 days after the outburst on November 1. TSpec is a cross-dispersed spectrograph with a $1\arcsec\times30\arcsec$ slit, which provided a resolving power of $\sim$2700 over 1--2.4\,$\mu$m \citep{2008SPIE.7014E..0XH}. For each run, dome flats were taken in the afternoon, and at night, 12P was observed together with two G-type stars \citep{2020AJ....160..130L} as both telluric correction standards and solar analogs. All nighttime frames were taken using in-slit nodding with a nodding distance of 15$\arcsec$ separation, and the slit was aligned along the parallactic angle. The TSpec spectra were extracted using the {\sc astro-plpy}\footnote{\label{fn:plpy}\url{https://github.com/RuiningZHAO/plpy}} package, where the wavelength calibration was performed using OH sky emission lines. The extracted spectra were then reduced to reflectance using a modified version of Spextool \citep{2004PASP..116..362C}.

One optical spectra of 12P was obtained with the Yunnan Faint Object Spectrograph and Camera (YFOSC) on the Lijiang 2.4-meter telescope at Gaomeigu site \citep{2019RAA....19..149W}. The observation was conducted on UT 2023 November 2, between the two TSpec observations. The G3 grism was used with the $1\farcs8\times9\farcm4$ slit, which provided a resolving power of $\sim320$ over 0.36--0.91\,$\mu$m. During the night, dome flats were taken first, followed by observations of science targets. In addition to 12P, two G2V stars and one flux standard nearby were observed. Wavelength calibration was performed using a helium-neon arc. The slit was aligned with the parallactic angle throughout the run. The data were reduced with {\sc astro-plpy}\footref{fn:plpy} package following standard procedure.

The observation log is presented in Table\,\ref{tab:obslog}, which lists all targets observed during the three runs, along with their exposure times and airmasses. Also tabulated are the heliocentric distance ($r_{\rm h}$), geocentric distance ($\Delta$), and phase angle ($\phi$) of 12P.


\begin{deluxetable*}{ccccccccc}
    \tablenum{1}
    \tablecaption{Log of observations.\label{tab:obslog}}
    \tablewidth{0pt}
    \tablehead{
        \colhead{Date of 2023} & \colhead{$\Delta t$\tablenotemark{a}} & \colhead{Tel./Inst.} & \colhead{Object} & \colhead{Exposure} & \colhead{Airmass} & \colhead{$r_{\rm h}$\tablenotemark{b}} & \colhead{$\Delta$\tablenotemark{c}} & \colhead{$\phi$\tablenotemark{d}}\\
        \colhead{(UT)} & \colhead{(d)} & \nocolhead{} & \nocolhead{} & \colhead{(s)} & \nocolhead{} & \colhead{(au)} & \colhead{(au)} & \colhead{($^\circ$)}
    }
    \decimals
    \startdata
    Oct. 9   & +4.07    & IRTF/SpeX   & 12P        & 18$\times$120 & 1.43--1.69 & 3.014    & 3.051    & 18.950\\
    $\cdots$ & $\cdots$ & $\cdots$    & HD 162209  & 14$\times$3   & 1.56--1.58 & $\cdots$ & $\cdots$ & $\cdots$\\
    $\cdots$ & $\cdots$ & $\cdots$    & BD+28 2993 & 8$\times$7    & 1.39--1.40 & $\cdots$ & $\cdots$ & $\cdots$\\
    \hline
    $\cdots$ & $\cdots$ & $\cdots$    & HD 234382  & 4$\times$30   & 1.36--1.37 & $\cdots$ & $\cdots$ & $\cdots$\\
    Nov. 2   & +0.71    & P200/TSpec  & 12P        & 8$\times$150  & 1.41--1.65 & 2.737    & 2.873    & 20.187\\
    $\cdots$ & $\cdots$ & $\cdots$    & HD 212816  & 8$\times$90   & 1.51--1.54 & $\cdots$ & $\cdots$ & $\cdots$\\
    \hline
    $\cdots$ & $\cdots$ & $\cdots$    & HD 152306  & 16            & 1.72       & $\cdots$ & $\cdots$ & $\cdots$\\
    $\cdots$ & $\cdots$ & $\cdots$    & HD 159222  & 10            & 1.43       & $\cdots$ & $\cdots$ & $\cdots$\\
    Nov. 2   & +1.11    & 2.4-m/YFOSC & 12P        & 8$\times$300  & 1.49--1.77 & 2.732    & 2.870    & 20.211\\
    $\cdots$ & $\cdots$ & $\cdots$    & HD 164595  & 16            & 1.97       & $\cdots$ & $\cdots$ & $\cdots$\\
    \hline
    $\cdots$ & $\cdots$ & $\cdots$    & HD 234382  & 16$\times$30  & 1.29--1.33 & $\cdots$ & $\cdots$ & $\cdots$\\
    Nov. 3   & +1.71    & P200/TSpec  & 12P        & 42$\times$150 & 1.22--1.98 & 2.726    & 2.864    & 20.245\\
    $\cdots$ & $\cdots$ & $\cdots$    & HD 177780  & 68$\times$30  & 1.54--2.03 & $\cdots$ & $\cdots$ & $\cdots$\\
    \enddata
    \tablenotetext{a}{Days after outburst. The outburst dates are taken from \citet{BAA2023}.}
    \tablenotetext{b}{Heliocentric distance}
    \tablenotetext{c}{Geocentric distance}
    \tablenotetext{d}{Phase angle}
\end{deluxetable*}

\section{Results} \label{sec:res}

\subsection{NIR Spectroscopy} \label{sec:res:nir}
\begin{figure}[ht]
    \includegraphics[width=0.8\textwidth]{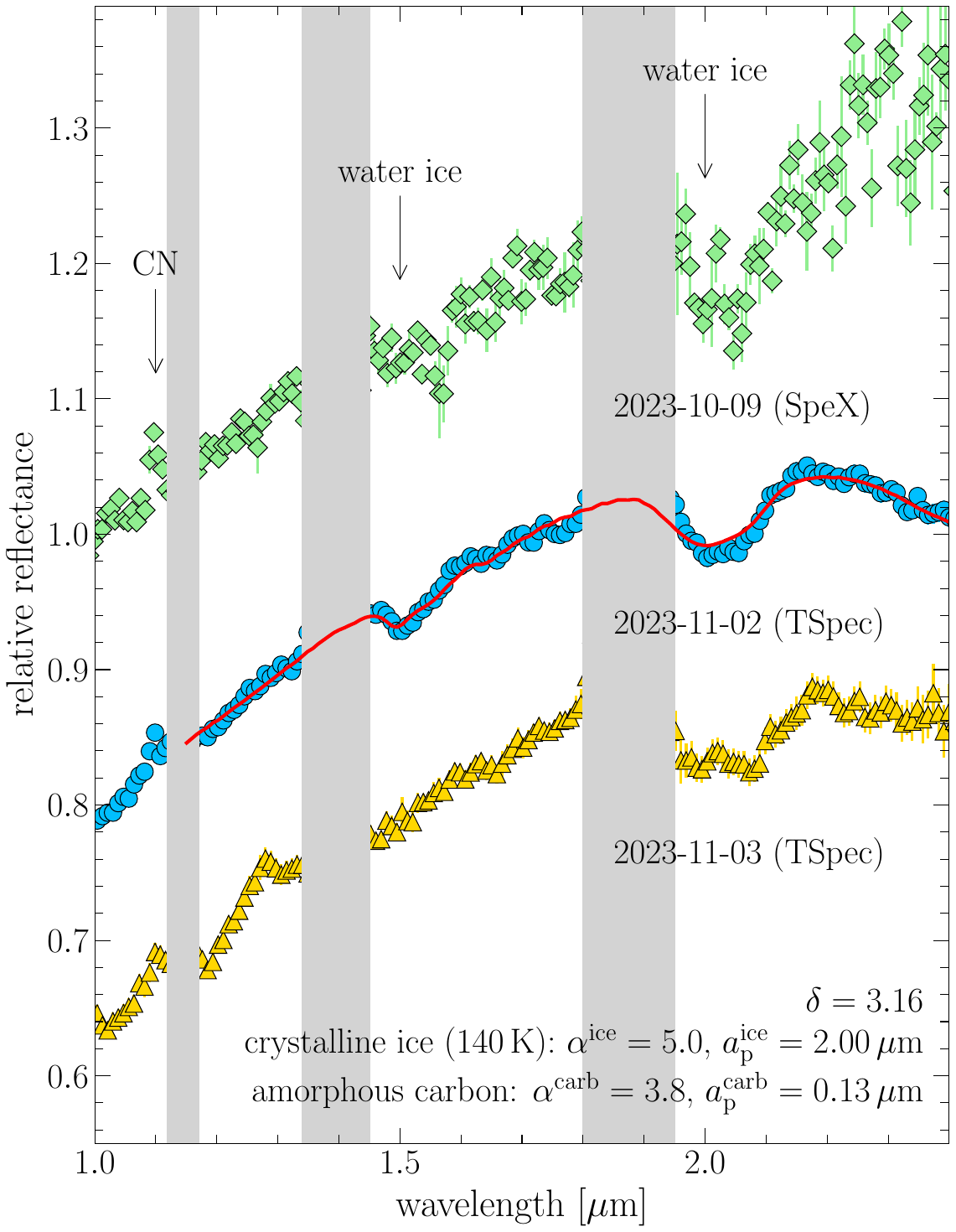}
    \caption{NIR spectra of 12P taken with SpeX and TSpec. The date and instrument used to obtain each spectrum are labeled below the respective spectrum. The two TSpec spectra are binned to a resolving power of $\sim$200. The red-system band of CN and the two water ice absorption bands are labeled. Spectral regions contaminated by strong telluric absorption are masked (gray shades). The ``emission'' band near 1.3\,${\rm \mu m}$ in the Nov. 3 spectrum is due to imperfect telluric correction (see \S\ref{sec:res:nir}). The best-fit model from \S\ref{sec:res:nir} is also shown (red line), with its parameters labeled at the bottom. \footnotesize\label{fig:nir}}
\end{figure}

Figure\,\ref{fig:nir} shows the NIR reflectance spectra of 12P, which consistently exhibit significant absorption bands superimposed on red continua. The absorption bands, centered at 1.5\,$\rm{\mu m}$ and 2.0\,$\rm{\mu m}$, are attributed to water ice grains. The spectral slopes over the 1.0--2.4\,${\rm \mu}$m range are 1.8$\pm$0.1\%, 2.0$\pm$0.1\%, and 2.3$\pm$0.1\% per 10$^3$\,${\rm\AA}$, respectively. The variation in spectral slope, however, may not be intrinsic, as different G-type stars are used to calculate the reflectances. 
As shown in Table\,\ref{tab:obslog}, the two G-type stars observed with SpeX (HD 162209 and BD+28 2993) were not included in the TSpec observations, making it infeasible to directly compare the slope differences between the October and November outbursts based solely on our data. For the two TSpec observations, the reported slopes, i.e., 2.0$\pm$0.1\% and 2.3$\pm$0.1\% per 1000\,${\rm\AA}$, are derived primarily using HD 234382 and HD 177780, respectively. Since both stars were observed on November 3 (see Table\,\ref{tab:obslog}), we find that using HD 177780 results in a slope that is 0.2\% per 1000\,${\rm\AA}$ larger than that obtained with HD 234382. Consequently, if the same G-type star were used, the slope difference between November 2 and 3 would reduce from 0.3\% to 0.1\% per 1000\,${\rm\AA}$. Given that this difference is within the range of random error, no intrinsic variation in spectral slope can be confirmed.

In addition, several emission features are also present. All three spectra consistently show a weak emission band at 1.1\,$\mu$m. It is attributed to the red-system band of the CN radical \citep{Shinnaka:2017}. Another ``emission'' band near 1.3\,${\rm \mu}$m in the November 3 spectrum arises from an imperfect removal of the telluric O$_2$ feature. Since 12P was setting during the observation, we had to observe the G-type standard star HD 177780 repeatedly across a wide range of airmass to match the changing elevation (see Table\,\ref{tab:obslog}). However, this introduced inconsistencies due to evolving atmospheric conditions that could not be fully corrected.


To constrain the detected water ice and the dust properties of 12P, we construct a spectral model to reproduce the spectrum obtained on November 2, which has the best signal-to-noise ratio (SNR) among the available observations. We consider an optically thin coma composed of two grain populations: water ice grains and refractory grains.
We do not expect water ice grains in cometary comae to be extremely pure, instead, they most likely have refractory (i.e., silicate and carbon) inclusions \citep[e.g., see][]{Greenberg:1999,Levasseur-Regourd:2004}. However, if the refractory inclusions only make up a tiny volume fraction (e.g., the ice coated on the refractory grains is much thicker than the refractory grains), they will not affect the optical properties of the ice grains. As will be shown later, our modeling reveals that the water ice grains are micrometer-sized while the refractory grains are submicrometer-sized, it is therefore justified that for the water ice grain population we will simply consider pure ice. At $r_h$\,$\sim$\,2.7--3\,au, micrometer-sized water ice grains attain an equilibrium temperature of $<$170\,K. 
In contrast, submicrometer-sized refractory grains with a thin ice coat attain a much higher temperature and are expected to lose their ice coats through sublimation. Therefore, we will consider a refractory population, represented by amorphous carbon or silicate dust. In principle, we could consider a porous aggregate of amorphous carbon and silicate for the refractory dust population. However, this will inevitably introduce several more model parameters. Given the limited amount of observational data, we prefer to simply approximate the refractory population by amorphous carbon. Nevertheless, we will show later that similar conclusions are achieved with silicate.  

The optical constants of water ice vary with its phase (amorphous or crystalline) and temperature. We use the optical constants provided by \citet{2008Icar..197..307M} for the amorphous ice at temperatures between 40 and 120\,K and for the crystalline phase at temperatures between 20 and 140\,K. The optical constants of amorphous carbon are taken from \citet{1991ApJ...377..526R}.

We assume that each of the two compositions follows a modified power-law differential size distribution over the range of $a_{\rm min}<a<a_{\rm max}$, i.e., 
\begin{equation}
    \frac{dn(a)}{da}\propto\left(1-\frac{a_{\rm min}}{a}\right)^{\gamma}\left(\frac{a_{\rm min}}{a}\right)^{\alpha}~~,
\end{equation}
where $\alpha$ is the power-law index, and $\gamma$ is related to the peak size $a_{\rm p}$ by $a_{\rm p}=(\gamma/\alpha+1)a_{\rm min}$ \citep{1986SSRv...43....1D}. In this work, we fix $a_{\rm min}$ and $a_{\rm max}$ to 10$^{-2}$ and 10$^4$\,$\mu$m, respectively. 

In cometary comae where the grains are sufficiently spaced, the optical depth is generally too low for multiple scattering to occur. This remains true even for outbursting comets, as shown by \cite{Lacerda:2012}, who measured a low optical depth of just 0.04 at 1.5$"$ from the nucleus of 17P/Holmes during its extraordinary 2007 outburst. They further inferred that the coma was optically thick only within 0.01$"$ of the nucleus; as such, the bulk of the coma remains optically thin. Under such optically thin conditions, the observed reflectance is directly proportional to the size-averaged angular scattering cross section, as described in Equation A1 of \citet{Bockelee-Morvan:2017}.

Using Mie Theory \citep{1983asls.book.....B}, we calculate for each composition the wavelength- and size-dependent scattering cross section $C_{\rm sca}(\lambda, a)$ and the asymmetry factor $g(\lambda,a)$, and then derive size-averaged angular scattering cross section of a given scattering angle $\theta$ by
\begin{equation}\label{eqn:csca}
    \left<C_{\rm sca}(\lambda)\right>=\int^{a_{\rm max}}_{a_{\rm min}} C_{\rm sca}(\lambda,a)\Phi_{\rm HG}[g(\lambda,a);\theta]\frac{dn(a)}{da} da~~,
\end{equation}
where $\Phi_{\rm HG}[g(\lambda, a);\theta]$ is the Henyey-Greenstein phase function \citep{1941ApJ....93...70H}. On November 2, the scattering angle is $\theta=\pi-\phi=159.8^{\circ}$. 

To represent a mixture of the two compositions, we introduce $\delta=M_{\rm carb}/M_{\rm ice}$, the refractory-to-ice mass ratio within the field-of-view (FoV), which defines the proportion of each component. The total reflectance is thus proportional to a linear combination of the mass-specific $\left<C_{\rm sca}(\lambda)\right>$ of the two compositions, i.e., 
\begin{equation}\label{eqn:mod}
    r(\lambda;\theta) \propto \frac{\left<C_{\rm sca}(\lambda)\right>_{\rm ice}}{\rho_{\rm ice}\left<V\right>_{\rm ice}}+\delta\frac{\left<C_{\rm sca}(\lambda)\right>_{\rm carb}}{\rho_{\rm carb}\left<V\right>_{\rm carb}}~~,
\end{equation}
where $\rho_{\rm carb}$ and $\rho_{\rm ice}$ are the mass densities taken from \citet{1991ApJ...377..526R} and \citet{2008Icar..197..307M}, with values of 1.85\,g\,cm$^{-3}$ and 0.9\,g\,cm$^{-3}$, respectively, and $\left<V\right>$ is the size-averaged volume.

Our best-fit model, shown as the red line in Figure\,\ref{fig:nir}, successfully reproduces the reflectance observed on November 2. The model consists of amorphous carbon grains and crystalline water ice grains at a temperature of 140\,K. The size distribution of the carbon grains follows a power-law index of 3.8 at larger sizes. It peaks at 0.13\,$\rm{\mu m}$, which is significantly smaller than the peak of the water ice grains at 2.00\,$\rm{\mu m}$. The size distribution of the water ice grains can be characterized by a power-law index of 5.0, where the population is dominated by small grains with a sharp drop in larger sizes. These ice grains contribute $\sim$ 24\% of the mass fraction (or $\sim$39\% of the volume fraction) in the coma.

To assess the robustness of the results, we replace the refractory material--amorphous carbon--with astronomical silicate \citep{1984ApJ...285...89D}. In this case, the best-fit size distribution parameters are $\alpha^{\rm ice}=5.1$, $a_{\rm p}^{\rm ice}=2.5$\,${\rm \mu}$m, $\alpha^{\rm sil}=4.2$, $a_{\rm p}^{\rm sil}=0.32$\,${\rm \mu}$m, with water ice grains accounting for 33\% of the mass (or 65\% of the volume). The two best-fit models yield similar size distributions, with water ice and refractory grains peaking at micrometer and sub-micrometer sizes, respectively. A major difference arises in the volume fraction of water ice grains, which can be readily explained by the higher scattering efficiency of silicate compared to amorphous carbon. Despite this difference, the consistency in the overall grain size distributions suggests that our grain model is robust in capturing the key properties of the water ice grains.


\subsubsection{Physical Properties of the Water Ice}

\begin{figure}[ht!]
    \includegraphics[width=0.8\textwidth]{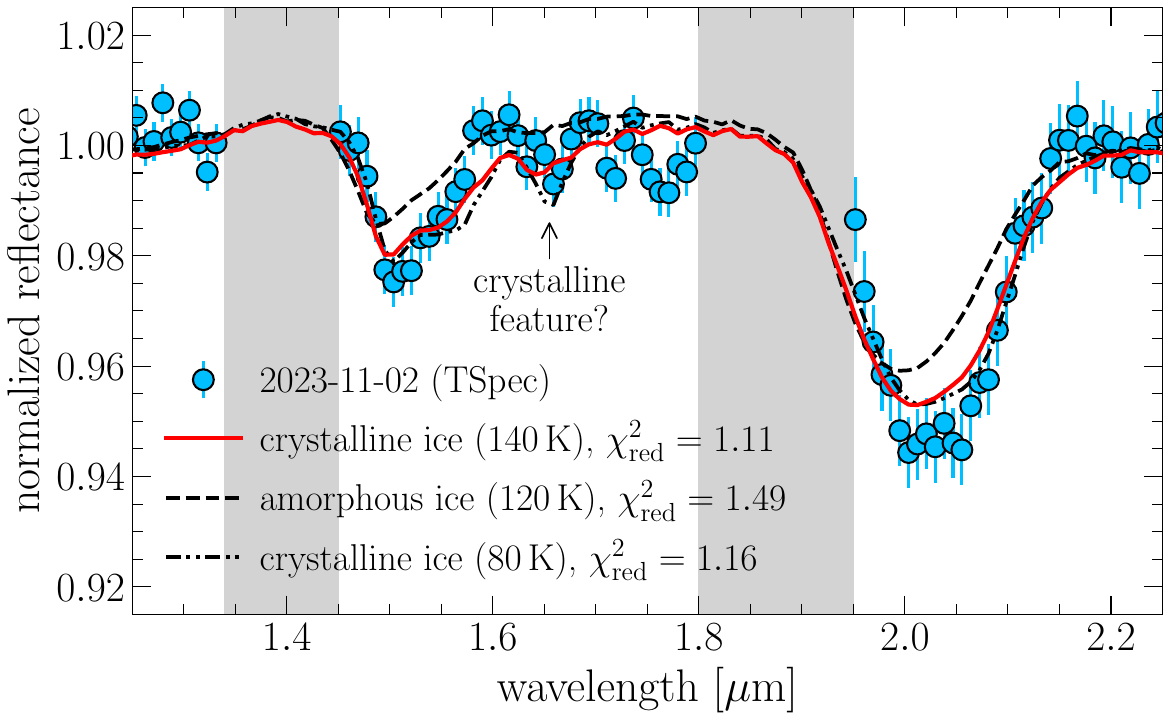}
    \caption{\footnotesize Water ice absorption bands taken on Nov. 2. The normalized spectrum (blue circles) is compared with the best-fit model (red line) and two alternative models in which the ice compositions are amorphous ice at 120\,K (dashed lines) and crystalline ice at 80\,K (dash-dotted lines). The error bars account for both random and systematic errors. Systematic errors in the {\it JHK} bands are estimated band-wise using the standard deviation within the continuum regions of 1.25--1.34, 1.70--1.80, and 2.15--2.45\,${\rm \mu}$m, respectively. \label{fig:mod1}}
\end{figure}

The high SNR of our data allows us to further investigate the detailed physical properties of the ice grains, including their phase and temperature. In Figure\,\ref{fig:mod1}, we present the continuum normalized absorption spectrum of 12P using the data taken on November 2. A narrow absorption feature is observed near 1.65\,${\rm \mu}$m, which could potentially indicate the presence of crystalline ice. However, besides the 1.65\,${\rm \mu}$m band, there are several narrow absorption features present in the 1.7--1.8\,${\rm \mu}$m range with a comparable strength to the former. This suggests that the 1.65\,${\rm \mu}$m band may not be real and could result from systematic errors, such as telluric absorption or instrumental fluctuations, and therefore cannot serve as a definitive indicator of crystallinity.

On the other hand, the band center and the spectral profile of the absorption feature at 1.5 and 2.0\,${\rm \mu}$m can also be used to distinguish amorphous ice from crystalline ice \citep{2008Icar..197..307M}. 
We derive two additional models by substituting only the optical constants of water ice in the best-fit model---one with crystalline ice at 80\,K and the other with amorphous ice at 120\,K---while keeping all other parameters unchanged.
As shown in Figure\,\ref{fig:mod1}, the two models with crystalline ice match the two broad absorption features significantly better compared to the amorphous ice model, as indicated by the lower reduced chi-square values (1.11 and 1.16 versus 1.49). The amorphous ice model produces absorption bands with centers shifted to shorter wavelengths and shallower depths relative to the observed data, resulting in a noticeable mismatch between the model and the spectral features. Between the two crystalline models, the one with 80\,K ice fits the 1.65\,${\rm \mu}$m feature well but exhibits a broader band width at 1.5\,${\rm \mu}$m, resulting in a higher overall chi-square value.

We note that the crystalline ice model at 140\,K provides the best fit but may not represent the optimal solution, as 140\,K is the highest temperature at which the optical constants were measured by \citet{2008Icar..197..307M}. Considering that sublimation can carry away a substantial amount of thermal energy, it is reasonable to adopt the blackbody equilibrium temperature $T\approx278\,{\rm K}\cdot(r_{\rm h}/{\rm au})^{-0.5}\approx170$\,K as the upper limit for the water ice. Thus, we infer that the water ice in the coma of 12P during the November outburst is predominantly, if not entirely, crystalline ice at temperatures $\simali$140--170\,K. 


\subsection{Optical Spectroscopy} \label{sec:res:opt}

\begin{figure}[ht]
    \includegraphics[width=0.8\textwidth]{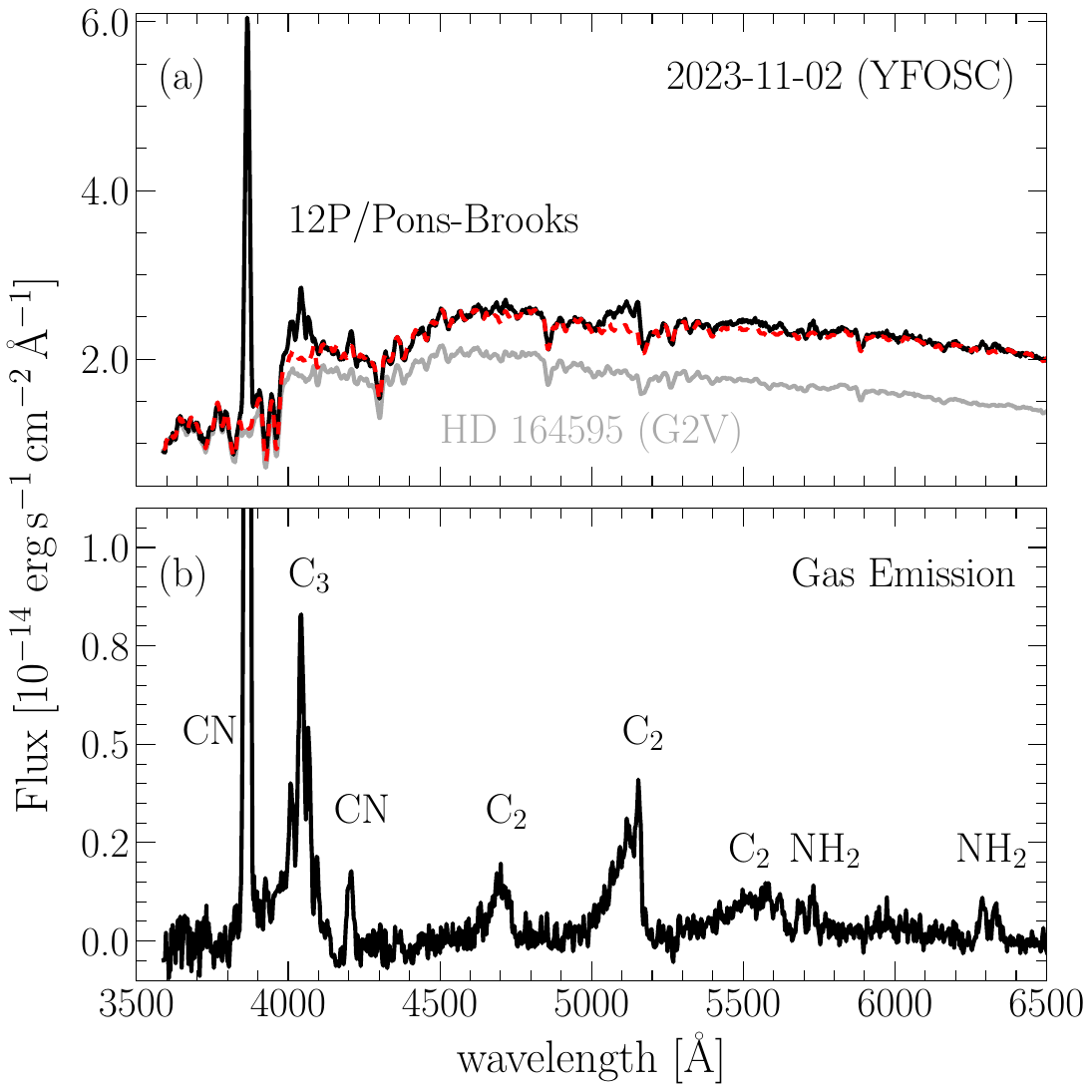}
    \caption{(a) YFOSC spectrum of 12P taken on 2023 Nov.\,2 (black line). Also shown are the spectrum of a G2V star (grey line) and the reproduced dust continuum (red dashed lines). (b) Emission component in the spectrum of 12P. Emission bands of CN, C$_3$, C$_2$, and NH$_2$ are labeled.\footnotesize\label{fig:opt}}
\end{figure}

We show in Figure\,\ref{fig:opt}a the optical spectrum of 12P,  extracted using a rectangular aperture of $1\farcs8\times28\farcs3$. The spectrum exhibits several gas emission bands superimposed on a dust continuum. To isolate the gas emissions, we model the continuum by multiplying the spectrum of HD 164595 (a G2V star) with a low-order polynomial and subtracting it from the comet spectrum. The resulting spectrum, consisting solely of gas emissions, is presented in Figure\,\ref{fig:opt}b, where emission bands of CN, C$_3$, C$_2$, and NH$_2$ are identified.

For quiescent comets, production rates of gaseous species and their photodissociation scale lengths are typically derived by fitting observed column density profiles with the Haser model \citep{1957BSRSL..43..740H}. 
For post-outburst comets, the original Haser model has been extended to include outburst-specific parameters, such as the quiescent and peak production rates, the expansion velocity of the ejected material, and the characteristic timescales of the outburst \citep{Opitom:2016,2024MNRAS.534.1816F}. However, the outburst model requires high-SNR spectra to sample column density profiles that extend sufficiently far from the nucleus to constrain the expansion pattern of the ejected material, typically $\rho>10^5$\,km \citep{Opitom:2016,2024MNRAS.534.1816F}. In our case, the aperture width of $28\farcs3$ (corresponding to $\rho\sim3\times10^4$\,km on either side of the nucleus) is insufficient for reliable parameter fitting. Therefore, we adopt a simplified approach, applying the original Haser model with fixed photodissociation scale lengths, and determining the production rates by matching the averaged column densities within our aperture to the model-predicted values. Assuming an expansion velocity of $v_{\rm exp}=0.85\,{\rm km\,s^{-1}}\cdot(r_{\rm h}/{\rm au})^{-0.5}$, and the $g$-factors and scale lengths from \citet{2012Icar..218..144C}, we find $Q_{\rm CN}=(1.82\pm0.01)\times10^{26}$\,s$^{-1}$, $Q_{\rm C_3}=(3.95\pm0.04)\times10^{25}$\,s$^{-1}$, $Q_{\rm C_2}=(1.42\pm0.02)\times10^{26}$\,s$^{-1}$, and $Q_{\rm NH_2}=(1.22\pm0.07)\times10^{26}$\,s$^{-1}$ for the November 1 outburst. 

To better interpret our results, it is important to distinguish between contributions from quiescent outgassing and those driven by the outburst, or at least to estimate their relative proportions. We use the large November 14 outburst as a reference. Applying an outburst model, \citet{2024MNRAS.534.1816F} derived quiescent production rates of $Q_{\rm CN} = 1.5 \times 10^{26}$\,s$^{-1}$ and $Q_{\rm C_3} = 5.7 \times 10^{24}$\,s$^{-1}$, and peak values during the outburst of $Q_{\rm CN} = 1.3 \times 10^{27}$s$^{-1}$ and $Q_{\rm C_3} = 1.2 \times 10^{26}$\,s$^{-1}$. In comparison, \citet{2023ATel16338....1J}, using a method similar to ours, reported production rates of $Q_{\rm CN} = (9.53 \pm 0.39) \times 10^{26}$\,s$^{-1}$ and $Q_{\rm C_3} = (2.69 \pm 0.24) \times 10^{26}$\,s$^{-1}$ about 1.08 days after the outburst. The latter are within a factor of two of the modeled peak rates, suggesting they reflect ongoing outburst-enhanced activity rather than quiescent emission. By analogy, we suggest that our measurements, obtained about 1.11 days after the November 1 outburst, are also representative of near-peak production rates.

We further calculate the production rate ratios of each species relative to CN: $\log\left(Q_{\rm C_3}/Q_{\rm CN}\right)=-0.66$, $\log\left(Q_{\rm C_2}/Q_{\rm CN}\right)=-0.11$, $\log\left(Q_{\rm NH_2}/Q_{\rm CN}\right)=-0.17$. Based on the classification system of \citet{2012Icar..218..144C}, 12P is ``typical'' in terms of C$_3$ abundance but ``depleted'' in C$_2$ during the November 1 outburst. This result is consistent with the findings of \citet{2024MNRAS.534.1816F}, derived using long-slit spectra of 12P taken in August and November 2023, although they classified 12P as `typical' in C$_2$ based on a different criterion.

The $Af\rho$ value, initially introduced by \citet{1984AJ.....89..579A}, has been widely used as an indicator of the dust production rate in comets. Assuming a $1/\rho$ brightness profile, we convert the rectangular aperture of the long-slit into an equivalent circular aperture of 2.55$\arcsec$ radius \citep{2019JOSS....4.1426M} and derive $Af\rho=(1.30\pm0.01)\times10^4$\,cm from the continuum flux at 5240\,${\rm \AA}$. Considering that 12P has brightened $\sim$3.7\,mag \citep[or a factor of $\sim$30;][]{BAA2023} and that its optical spectrum is dominated by continuum flux, the $Af\rho$ value at quiescent state should be at least 30 times less, on the order of $\sim$10$^2$--10$^3$\,cm. 
This places 12P within the range of $Af\rho$ values observed in both the five Halley-type comets studied by  \citet{Fink:2009} and the 42 Jupiter-family comets reported by \citet{Gillan:2024}.


\section{Discussion} \label{sec:dis}

\subsection{Water ice grains} \label{sec:ice}

\begin{figure}[ht!]
    \includegraphics[width=0.8\textwidth]{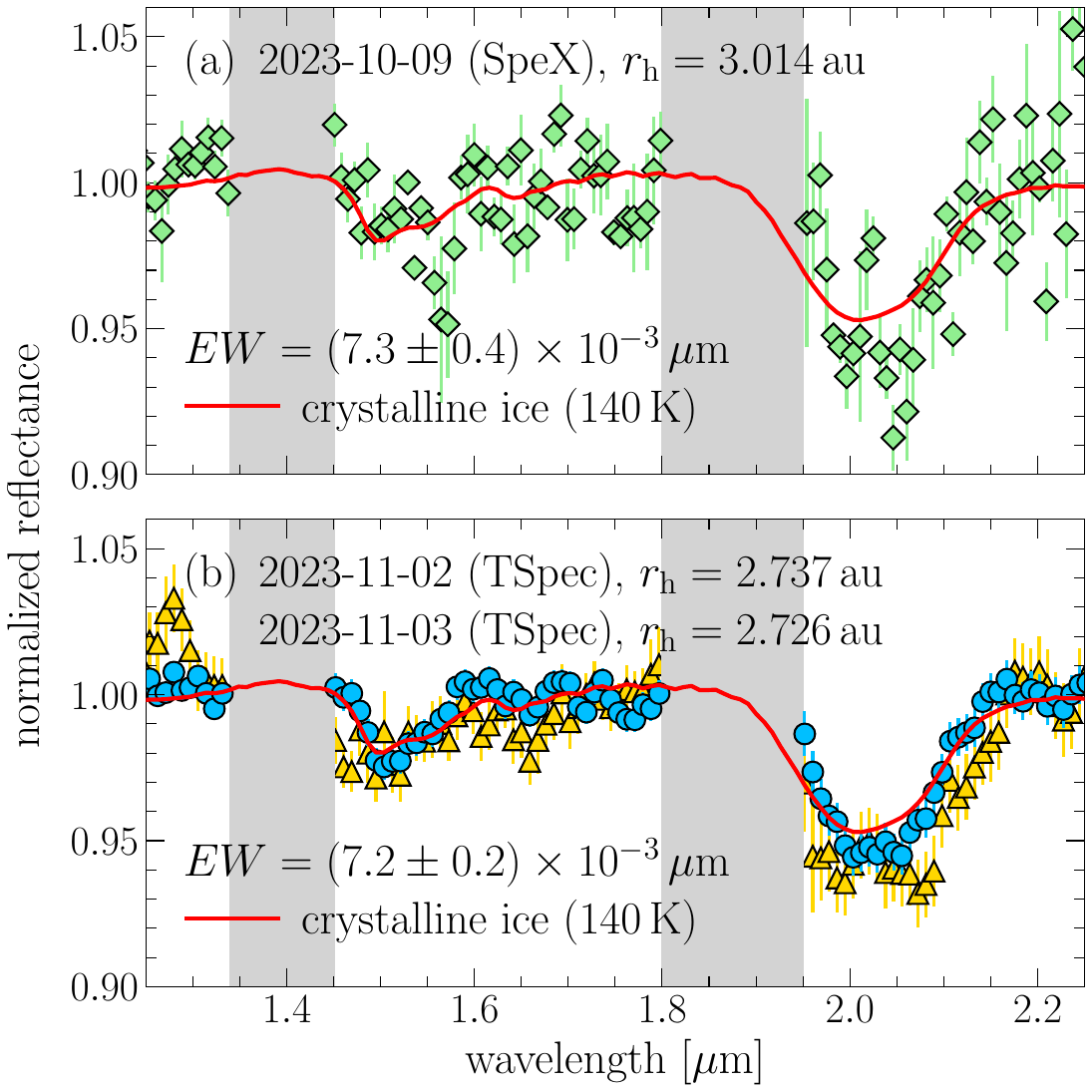}
    \caption{\footnotesize Comparison of water ice absorption bands during the two outbursts. The strengths of the 2\,${\rm \mu}$m absorption band are quantified using their equivalent widths ($EW$). Also shown are the heliocentric distance ($r_{\rm h}$) of the comet and the best-fit dust model in \S\ref{sec:res:nir} (red line). The fluctuations in the Nov. 3 spectrum (yellow triangle) including an ``emission'' band near 1.3\,${\rm\mu}$m is due to imperfect telluric correction (see \S\ref{sec:res:nir}). \label{fig:mod2}}
\end{figure}

The outburst in November brightened 12P by a factor of $\sim$30 \citep{BAA2023}, suggesting that the coma was primarily composed of freshly ejected grains. By the time of observation on November 2, water ice grains had persisted for 0.71 days ($\sim6\times10^4$\,s; see Table \ref{tab:obslog}). If the grains were predominantly pure with only a tiny amount of contamination (as assumed in \S\ref{sec:res:nir}), they would have undergone little evolution during this period, and their size distribution would remain nearly unchanged since the outburst. This is consistent with our observations, as the spectra obtained on the consecutive night appear sufficiently similar (see Figure\,\ref{fig:mod2}). Alternatively, the grains could have been impure and evolved into the observed distribution over that time. However, micrometer-sized water ice grains, dominant in the best-fit size distribution, can survive only $\sim3\times10^3$\,s at similar heliocentric distances if they contain impurities \citep{2006Icar..180..473B}. This would have led to noticeable spectral changes by the following day, which contradicts our observations on November 3. We therefore conclude that the water ice grains in 12P’s coma were predominantly pure, with minimal contamination.



Theoretical calculations indicate that water ice can only be remotely detected at distances beyond $\sim$2.5\,au from the Sun. At closer distances, icy grains evaporate rather rapidly, limiting the water ice halo to just a few hundred kilometers, smaller than the resolution achievable by ground-based telescopes \citep{1981Icar...47..342H}. 
Indeed, remote detections of water ice within $\sim$2.5\,au are exceedingly rare, limited to cases of direct  {\it in situ} measurements \citep{2014Icar..238..191P} and unusual cometary activity, such as the outburst of 17P/Holmes \citep{2009AJ....137.4538Y}.
The October and November outbursts of 12P occurred at $\sim$3.0 and $\sim$2.7\,au, respectively, with the second being very close to the theoretical limit. If the dust properties (e.g. grain size or composition) and abundance had remained constant during this period, the absorption bands would have weakened during the November outburst due to the rapid evaporation of icy grains. However, the equivalent widths of the 2\,${\rm \mu}$m absorption band measured on October 9 and November 2 are $(7.3\pm0.4)\times10^{-3}$ and $(7.2\pm0.2)\times10^{-3}$\,${\rm \mu}$m, respectively (see Figure\,\ref{fig:mod2}), showing no significant decrease in band depth. This suggests that some factors counteracted the expected weakening, such as a size distribution slightly shifting toward larger grains, or a higher replenishment rate of icy grains.



\subsection{Characterization of the November outburst} \label{sec:outburst}

We quantitatively characterize the November outburst by integrating the best-fit dust properties with the existing knowledge of 12P, and briefly discuss the underlying outburst mechanism.

{\it Total grain mass loss, $M_{\rm gra}$.} 
The total scattering cross section is related to the $Af\rho$ value by $C_{\rm sca}^{\rm total}=(\pi\rho/4\pi)Af\rho$ \citep{2012Icar..221..721F}. Taking the ice-mass-specific scattering cross section from Equation (\ref{eqn:mod}), the total masses of water ice and amorphous carbon grains in the coma are calculated as
\begin{equation}
    M_{\rm ice}=(\pi\rho/4\pi)Af\rho\left/\left[\frac{\left<C_{\rm sca}(\lambda)\right>_{\rm ice}}{\rho_{\rm ice}\left<V\right>_{\rm ice}}+\delta\frac{\left<C_{\rm sca}(\lambda)\right>_{\rm carb}}{\rho_{\rm carb}\left<V\right>_{\rm carb}}\right]\right.~~,
\end{equation}
and 
\begin{equation}
    M_{\rm carb}=M_{\rm ice}\delta~~.
\end{equation}
Given that our model accurately reproduces the observed NIR reflectance, we consider it a reliable representation of the dust's spectral behavior. This allows us to evaluate the above equations at any wavelength within the range defined by the available optical constants. However, our NIR spectra of the comet are not flux calibrated, preventing a direct derivation of NIR $Af\rho$. Furthermore, the post-outburst surface brightness distribution is possibly not $1/\rho$ and likely changes with time. But, under the assumption that the grain properties are constant with aperture size and time, we can estimate the NIR $Af\rho$ at the time and aperture size of the optical spectrum.
After scaling the NIR data to form a continuous spectrum, $Af\rho$ depends only on reflectance. Here, we focus on 1.15\,$\rm{\mu}$m, the shortest NIR wavelength closest to the optical regime. We find that the reflectance at 1.15\,$\rm{\mu}$m is approximately 40\% higher than that at 5240\,\AA. Accordingly, we estimate $Af\rho \sim 1.8 \times 10^4$ cm at 1.15\,$\rm{\mu}$m.
Substituting this value, and evaluating $\left<C_{\rm sca}(\lambda)\right>_{\rm ice}$ and $\left<C_{\rm sca}(\lambda)\right>_{\rm carb}$ at 1.15\,${\rm \mu}$m, we find $M_{\rm carb}\sim3.5\times10^8$\,kg, $M_{\rm ice}\sim1.1\times10^8$\,kg, and a total grain mass of $M_{\rm gra}=M_{\rm carb}+M_{\rm ice}\sim4.6\times10^8$\,kg. This should be considered a lower limit, as the slit might not fully encompass the ejecta, and very large grains, which contribute significantly to the coma mass, scatter poorly at short wavelengths.
Since the coma was dominated by grains produced during the outburst, $M_{\rm gra}\gtrsim4.6\times10^8$\,kg also represents the total mass loss from the outburst, and it is roughly consistent with a numerical model in the case of ice-controlled sublimation with 10\% active surface area \citep{Gritsevich:2025}.

\citet{Ye:2020} determined a nuclear radius of $\sim$17\,km based on early recovery imaging. However, their measurement did not account for contamination from dust and gas emissions, making it an upper-limit. Meanwhile, Li et al. (submitted) derived a lower limit of 2.84\,km based on the peak active area. Assuming a bulk mass density of 500\,kg\,m$^{-3}$, we find a nuclear mass of $5\times10^{13}$--$1\times10^{16}$\,kg. Therefore, the total mass loss from the November outburst accounts for $\gtrsim5\times10^{-8}$--$\gtrsim1\times10^{-5}$ of the nuclear mass. The upper limit of $\gtrsim1\times10^{-5}$ is comparable to the outbursts of 332P/Ikeya--Murakam, P/2010 H2 (Vales), and 15P/Finlay, while at least 3--4 orders of magnitude lower than 17P/Holmes \citep{Jewitt:2020,Ishiguro:2016}.

{\it Kinetic energy, $E_{\rm k}$.} 
For each dust composition, the total kinetic energy is calculated as
\begin{equation}
    E_{\rm k}=\frac{N}{2}\left(\frac{4}{3}\pi\rho_{\rm d}\right)\left<a^3V^2\right>~~,
\end{equation}
where $N$ is the total number of grains, $\rho_{\rm d}$ is the grain mass density, and $V$ is the grain expansion velocity. The angle brackets denote an average over the grain size distribution. Taking $\left<C_{\rm sca}(\lambda)\right>$ defined by Equation\,(\ref{eqn:csca}), the total number is calculated as $N=C_{\rm sca}^{\rm total}/\left<C_{\rm sca}(\lambda)\right>=(\pi\rho/4\pi)Af\rho/\left<C_{\rm sca}(\lambda)\right>$. Taking the best-fit size distributions and assuming a velocity-size relation of $V(a)=V_0(a/a_0)^{-0.5}$ with $V_0\sim0.33$\,km\,s$^{-1}$ and $a_0\sim1.5$\,${\rm\mu}$m from imaging \citep{Ryske:2023,Manzini:2023}, the total kinetic energy is derived to be $E_{\rm k}^{\rm car}\sim2.3\times10^{12}$\,J and $E_{\rm k}^{\rm ice}\sim0.9\times10^{12}$\,J, respectively. The corresponding specific kinetic energy is  $E_{\rm k}/M\sim8\times10^3$\,J\,kg$^{-1}$ for both compositions. 

The total kinetic energy of 12P is comparable to 15P/Finlay and 332P/Ikeya--Murakami, about an order of magnitude higher than P/2010 H2 (Vales), yet significantly lower than that of 17P/Holmes during its 2007 outburst \citep{Ishiguro:2016}. However, the specific kinetic energies of these comets are comparable, suggesting a similar triggering mechanism. For 17P/Holmes and 332P/Ikeya--Murakami, the outbursts were likely driven by the crystallization of amorphous water ice, which can provide a sufficient energy supply of $9\times10^4$\,J\,kg$^{-1}$ \citep{Li:2011,Ishiguro:2014}, although other mechanisms, such as rotational breakup of the nuclei, could not be ruled out \citep{Li:2015,Ishiguro:2016}. A similar mechanism was proposed for 1P/Halley to explain an eruption at large $r_{\rm h}$ \citep{Prialnik:1992b}. Given that 12P is a Halley-type comet, it is reasonable to propose the same mechanism for its outburst. The rotation period of 12P, estimated at $\sim$57 hours \citep{Knight:2024}, does not support the rotational breakup scenario. While other mechanisms cannot be entirely ruled out due to limited evidence, the crystallization of amorphous water ice remains the most plausible explanation.


{\it Refractory-to-ice mass ratio, $\delta_{\rm lost}$.}
In previous section (see \S\,\ref{sec:res:nir}), our best-fit model yields a refractory-to-ice mass ratio of $\delta\sim3.2$ in the coma. However, since gas mass loss is not included in the model, this value represents only an upper limit on the refractory-to-ice ratio in the material lost from the nucleus, i.e., $\delta_{\rm lost} < 3.2$.

To constrain the lower limit, we estimate the gas mass loss and incorporate it into $M_{\rm ice}$. Taking $Q_{\rm CN}/Q_{\rm OH}=6.76\times10^{-3}$ \citep{2023ATel16338....1J}, we find $Q_{\rm OH}\sim2.69\times10^{28}$\,s$^{-1}$. Assuming $Q_{\rm OH}/Q_{\rm H_2O}\sim0.9$ \citep{Fink:1990} and that water constitutes $\sim$80\% of the total gas mass, we find $\dot{M}_{\rm gas}\sim10^3$\,kg\,s$^{-1}$ at the peak of the outburst. Taking half of this value as an upper limit for the average gas mass loss rate during the outburst and adopting a duration of $\sim2\times10^5$\,s, as estimated for the November 14 outburst by \citet{2024MNRAS.534.1816F}, we find $M_{\rm gas}<10^8$\,kg. Thus, a lower limit is given by $\delta_{\rm lost}=M_{\rm carb}/(M_{\rm ice}+M_{\rm gas})>1.7$, leading to the final range of $1.7<\delta_{\rm lost}<3.2$.

Assuming reasonable densities for the dust materials, \citet{Patzold:2019} derived a refractory-to-ice mass ratio of 3--7 in the lost materials based on the measured bulk density of 67P/Churyumov-Gerasimenko’s nucleus.
Values $<3$ were not favored, as they would imply a porous icy nucleus with minimal volume of dust materials \citep{Patzold:2019}. Independent measurements further support the lower bound ($\gtrsim3$), including bulk density estimates of ejected dust particles \citep{Fulle:2017}, mass transfer analyses in perihelion chunks \citep{Fulle:2019}, and dielectric permittivity measurements \citep{Herique:2016}. However, \citet{Choukroun:2020} obtained a significantly lower range of 0.2--3 by adding ice to dry porous dust particles collected by {\it Rosetta}/COSIMA until their density matched that of 67P’s nucleus, suggesting that a definitive value has yet to be established. Similarly, during {\it Giotto}’s encounter with 1P/Halley, a dust-to-gas mass ratio of 1.3--3 was inferred as indicative of the nucleus’s refractory-to-ice ratio, though later revisions based on updated dust size distributions suggested larger values \citep{McDonnell:1991,Fulle:2000}. For 12P, the derived range of 1.7--3.2 aligns with previous lower-end estimates, potentially indicating a less evolved nucleus or an outburst region enriched in icy material compared to the bulk nucleus.


\section{Summary} \label{sec:sum}

Dust grains and gas in the coma of 12P/Pons-Brooks have been detected and characterized through its optical and NIR spectra obtained during two outbursts on October 5 and November 1, 2023. Our main findings are as follows:
\begin{enumerate}
    \item The NIR reflectance of 12P can be explained by sub-micrometer-sized amorphous carbon grains and micrometer-sized crystalline water ice grains at 140--170\,K. The water ice grains have a steeper slope at the large-size end of the size distribution, and contribute $\sim$24\% of the mass fraction (or $\sim$39\% of the volume fraction) in the coma.
    \item The optical spectrum during the November outburst revealed emission bands of CN, C$_3$, C$_2$, and NH$_2$. 12P is ``typical'' in terms of C$_3$ abundance but near the limit of C$_2$-depletion. The $Af\rho$ value at quiescent state is on the order of $\sim$10$^2$--10$^3\,$cm, within the span observed for Halley-type and Jupiter-family comets at similar $r_{\rm h}$.
    \item No significant weakening is observed in the depth of the water ice absorption bands from October to November, despite stronger sublimation expected at smaller heliocentric distances. Possible explanations include a size distribution slightly shifting toward larger grains, or a higher replenishment rate of icy grains. 
    \item The specific kinetic energy of the November outburst, $E_{\rm k}/M\sim8\times10^3$\,J\,kg$^{-1}$, suggests a triggering mechanism similar to 332P/Ikeya--Murakami and 17P/Holmes, likely the crystallization of amorphous water ice, though other mechanisms remain possible due to evidence.
    \item A refractory-to-ice ratio of $1.7<\delta_{\rm lost}<3.2$ is derived based on the total mass loss of dust and gas mass, aligning with the lower-end estimates for 67P/Churyumov-Gerasimenko and 1P/Halley. This suggests a less evolved nucleus or an outburst region enriched in icy materials compared to the bulk nucleus.
\end{enumerate}

\begin{acknowledgments}
    We thank the anonymous referee for their valuable comments and suggestions which improved the quality and presentation of this work. This work is sponsored (in part) by the Chinese Academy of Sciences (CAS), through a grant to the CAS South America Center for Astronomy (CASSACA). This research uses data obtained through the Telescope Access Program (TAP), which has been funded by the TAP association, including the Center for Astronomical Mega-Science CAS (CAMS), XMU, PKU, THU, USTC, NJU, YNU, and SYSU. We acknowledge the support of Dr. Jianguo Wang and the staff of the Lijiang 2.4m telescope. Funding for the telescope has been provided by Chinese Academy of Sciences and the People's Government of Yunnan Province. R.N.Z. and Y.H. acknowledge support from the National Key Basic R\&D Program of China via 2023YFA1608303. R.N.Z. and B. Y.  are visiting astronomers at the Infrared Telescope Facility, which is operated by the University of Hawaii under contract 80HQTR24DA010 with the National Aeronautics and Space Administration. B.Y. was supported by the China-Chile joint research funding CCJRF2209. M.S.P.K. and S.P. were supported by the National Aeronautics and Space Administration (USA) Solar System Observations program (grant no.\ 80NSSC20K0673). J.F.L. acknowledges the support from the New Cornerstone Science Foundation through the New Cornerstone Investigator Program and the XPLORER PRIZE. 
\end{acknowledgments}

%

\vspace{5mm}
\facilities{IRTF (SpeX), Hale (TSpec), YAO:2.4m (YFOSC)}


\software{astropy \citep{2022ApJ...935..167A}, 
          astro-wcpy \citep{2023ascl.soft11001Z}, 
          sbpy \citep{2019JOSS....4.1426M}
          }



\bibliography{sample631}{}
\bibliographystyle{aasjournal}



\end{document}